\documentclass[%
reprint,
superscriptaddress,
amsmath,amssymb,
]{revtex4-1}

\usepackage{color}
\usepackage{graphicx}
\usepackage{dcolumn}
\usepackage{bm}

\begin {document}
\preprint{APS/123-QED}

\title{Distributional Ergodicity in Stored-Energy-Driven L\'evy Flights}

\author{Takuma Akimoto}
\email{akimoto@z8.keio.jp}
\affiliation{%
  Department of Mechanical Engineering, Keio University, Yokohama, 223-8522, Japan
}%

\author{Tomoshige Miyaguchi}
\email{tmiyaguchi@naruto-u.ac.jp}
\affiliation{%
  Department of Mathematics Education, Naruto University of Education, Tokushima 772-8502, Japan
}%


\date{\today}

\begin{abstract}

  We study a class of random walk, the stored-energy-driven L\'evy flight
  (SEDLF), whose jump length is determined by a stored energy during a
  trapped state.
The SEDLF is a continuous-time random walk with jump lengths being coupled
with the trapping times.  It is analytically shown that the
ensemble-averaged mean square displacements exhibit subdiffusion as well as
superdiffusion, depending on the coupling parameter.
We find that time-averaged mean square displacements increase linearly with
time and the diffusion coefficients are intrinsically random, a
manifestation of {\it distributional ergodicity}.  The diffusion
coefficient shows aging in subdiffusive regime, whereas it increases with
the measurement time in superdiffusive regime.
\end{abstract}

\maketitle


\section {Introduction}

Single particle tracking experiments in biological systems often show that diffusion 
 is not normal but rather anomalous \cite{Caspi2000, Golding2006, Graneli2006, Weigel2011, Jeon2011, Weber2012}; 
that is, the mean square displacement (MSD) does not grow linearly with time but follows 
a power-law scaling
\begin{equation}
\langle x_t^2 \rangle \propto t^\beta \quad(\beta\ne 1). 
\end{equation} 
 Because anomalous diffusions including subdiffusion ($\beta<1$) as well as
  superdiffusion ($\beta>1$) are ubiquitously observed in many biological experiments, 
   anomalous diffusion is believed to play significant roles in cell biology such as
  gene regulation \cite{Zaid2009} and active transports \cite{Caspi2000, Weber2012}.
However,   the underlying physical mechanisms remain controversial. 
 
 To understand the underlying mechanisms of these anomalous diffusions, 
 phenomenological models such as 
 continuous-time random walk (CTRW), L\'evy walk and flight, and other 
 stochastic models of anomalous diffusion have been intensively studied 
 \cite{metzler00, saxton07, Lubelski2008, He2008, Jeon2011, Miyaguchi2011a, Meroz2013}.
Among these models, CTRW shows a prominent feature called 
  {\it distributional ergodicity} \cite{Lubelski2008, He2008, Miyaguchi2011a, Miyaguchi2011}; that is, the time average of an observable
  converges to a random variable, i.e., {\it convergence in distribution}, but it does not
  coincide with the ensemble average as in the ordinary sense of ergodicity. 
   It is considered that this
  distributional behavior of time-averaged observables in CTRW is
  related to large fluctuations of transport coefficients in single
  particle tracking experiments \cite{Golding2006, Graneli2006, Weigel2011, Jeon2011}. 
  It is known that such distributional behavior is universal in {\it infinite ergodic theory} \cite{Aaronson1997, Akimoto2010}, 
  where ergodicity is satisfied with respect to an infinite (non-normalizable) invariant measure.  
  This is a completely different feature from other stochastic models of subdiffusion. 

  While uncoupled CTRWs, in which trapping time and jump length are
  mutually independent, are extensively studied, effects of a coupling
  between them become physically important for nonthermal systems such as
  cells \cite{Caspi2000, Weber2012}. In such nonthermal systems, a particle
  in a trapped state would not be simply frozen, but rather it would be
  storing a sort of energy for the next jump.  Thus, a random walk driven
  by an stored energy during a trapped state is essential in such
  nonthermal systems, and it will also be important in complex systems such

  as 
  finance \cite{Meerschaert2006} and earthquakes \cite{Helmstetter2002}.
  
  As a prototype model of such nonthermal random walks, we study a CTRW
  with jump lengths correlated with trapping times \cite{Klafter1987,
    Magdziarz2012, Liu2013}, which we refer to as the stored-energy-driven
  L\'evy flight (SEDLF).  The SEDLF exhibits a whole spectrum
    of diffusion: sub-, normal-, and super-diffusion, depending on a
    parameter $\gamma$, which characterizes the coupling strength between
    jump length and trapping time.  Here, we show a novel type of
    distributional ergodicity. In particular,  time-averaged observables
  such as the time-averaged MSDs (TAMSDs) are intrinsically random even
  when the measurement time goes to infinity.

\section {Model}
\begin{figure}
  \includegraphics[width=.9\linewidth, angle=0]{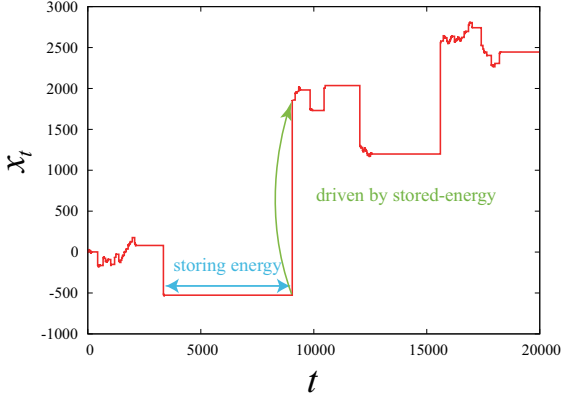}
  \caption{ A trajectory of SEDLF ($\alpha=0.7$ and
      $\gamma=0.9$). A big jump occurs when a random walker is trapped for
      a long time.}
  \label{traf_alpha=0.7}
\end{figure}

  The SEDLF is based on  CTRW with a non-separable
  joint probability of trapping time and jump length. In general,  CTRW is
  defined through the joint probability density function (PDF) $\psi (x, t)$,
  where $\psi (x, t)dx dt$ is the probability that a random walker jumps with
  length $[x, x+dx)$ just after it is trapped for period $[t, t+dt)$ since its previous
  jump \cite{Shlesinger1982, bouchaud90}. In particular, the separable
case $\psi(x,t) = w(t) l(x)$, in which the jump length and the trapping time
are mutually independent, has been extensively studied \cite{Scher1975,
  metzler00, Lubelski2008, He2008, Miyaguchi2011a}.  Here, we consider a
non-separable case defined by
\begin{equation}
  \label{e.joint-prob}
  \psi(x,t) = w(t) \frac {\delta (x-t^{\gamma}) + \delta (x+t^{\gamma})}{2},
\end{equation}
where $w(t)$ is the PDF of trapping times and $\gamma \in [0,1]$ is a coupling strength. 
Note that a random walker undergoes a long trapped state before it performs a long jump 
(Fig.~\ref{traf_alpha=0.7}).
In addition, we assume that the PDF of trapping times follows a power
law:
\begin{equation}
  \label{e.pdf-trap-jump}
  w(t) \simeq \frac {c_0}{t^{1+\alpha}},
\end{equation}
as $t\to \infty$. Here, $\alpha \in (0,1)$ is the stable index, a constant
$c_0$ is defined by $c_0 = {c}/{|\Gamma(-\alpha)|}$ with a scale factor
$c$. 

For $\gamma=0$, the SEDLF is just a separable CTRW with jumps only to the
nearest neighbor sites. On the other hand, for $\gamma>0$, the PDF of jump
  follows a power law:
  \begin{equation}
    \label{e.pdf-jump-length}
    l(x) = \int_{0}^{\infty} \psi (x,t) dt  =
    \frac {|x|^{\frac {1}{\gamma}-1}}{2\gamma}w\left(|x|^{\frac
      {1}{\gamma}}\right)
    \simeq \frac {c_0}{2\gamma}\frac {1}{|x|^{1+ {\alpha} / {\gamma}}}.
  \end{equation}
  Thus, the mean jump length diverges for $\gamma \geq \alpha$.
  Note that the L\'evy flight also has a power law distribution of jump
  length, which causes a divergence in the MSD. By contrast, the MSD of the
  SEDLF is finite with the aid of the coupling between jump lengths and trapping
  times as shown below. This property makes the SEDLF a physically more
  coherent model than L\'evy flight.

\section {Theory}
Generalized master equations for CTRWs obtained in
  \cite{Shlesinger1982} can be utilized for our model. In general, the
spacial distribution $P(x,t)$ of CTRWs with initial distribution
$P_0 (x)$ at time zero satisfies the following equations:
\begin{align}
  \label{e.recursion}
  P(x,t) &= \int_{0}^{t} dt' \Psi (t-t') Q(x,t') + \Psi_0(t) P_0(x),
  \\[0.1cm]
  Q (x,t) &=
  \int_{-\infty}^{\infty} dx' \int_{0}^{t} dt' \psi (x', t') Q (x-x', t-t')
  \nonumber\\[0.0cm]
  &\quad+ \int_{-\infty}^{\infty} dx' \psi_0 (x', t) P_0(x-x'),
\end{align}
where $Q(x,t) dt dx$ is the probability of a random walker reaching an
interval $[x,x+dx)$ just in a period $[t, t+dt)$, $\Psi (t)$ is the
probability of being trapped for longer than time $t$:
\begin{equation}
  \label{e.psi-cctrw}
  \Psi (t)
  =
  1 - \int_{-\infty}^{\infty}dx' \int_{0}^{t} \psi (x', t') dt'
  =
  1 - \int_{0}^{t} w (t') dt',
\end{equation} 
$\psi_0(x,t)$ is the joint PDF for the first jump, and $\Psi_0(t)$ is
the probability that the first jump does not occur until time $t$.
Fourier-Laplace transform with respect to space and time ($x\rightarrow k$
and $t\rightarrow s$, respectively),  defined by
\begin{equation} 
{\hat{P}}(k,s) \equiv \int_{-\infty}^\infty dx \int_0^\infty dt P(x,t) e^{ikx} e^{-st} , 
\end{equation}
 gives
  \begin{eqnarray}
    {\hat{P}}(k,s) 
    &=& \frac{\hat{P}_0(k)}{1-{\hat{\psi}}(k,s)} \frac{1-\hat{w}_0(s) + \hat{\varphi}_0(k,s)}{s},
    \label{masterl_eq}
  \end{eqnarray}
where $\hat{\varphi}_0(k,s) = [1-\hat{w}(s)]\hat{\psi}_0(k,s) - [1-\hat{w}_0(s)]\hat{\psi}(k,s)$.

In the case of the SEDLF, 
we obtain ${\hat{\psi}} (k,s)$ from Eq.~(\ref{e.joint-prob}) as follows
\begin{equation}
  \label{e.joint-prob-fourier-laplace}
  {\hat{\psi}} (k,s) =
  \int_{0}^{\infty} e^{-st} \cos \left( k t^{\gamma}\right) w(t) dt.
\end{equation}
Note that ${\hat{\psi}} (0,s) = \hat{w} (s)$, and the asymptotic
behavior of the Laplace transform of $w(t)$ [Eq.~(\ref{e.pdf-trap-jump})]
is given by
\begin{equation}
  \label{e.trap-pdf-laplace}
  1- \hat{w}(s) \simeq c s^{\alpha}\quad (s\rightarrow 0).
\end{equation}
We  assume that the initial distribution $P_0(x)$ is the delta function, 
$P_0(x)=\delta(x)$, and $w(t)=w_0(t)$ (ordinary renewal process \cite{Cox}). 
As a result, we have the following generalized master equation in the Fourier and Laplace space:
\begin{equation}
  \label{e.renewal-cctrw}
  {\hat{P}}(k,s) =
  \frac {1}{s}
  \frac{1-\hat{w}(s)}{1-{\hat{\psi}}(k,s)}, 
\end{equation}
where ${\hat{\psi}}(k,s)$ and $\hat{w}(s)$ are given by
Eqs.~(\ref{e.joint-prob-fourier-laplace}) and (\ref{e.trap-pdf-laplace}).

\begin{figure}
  \includegraphics[width=.9\linewidth, angle=0]{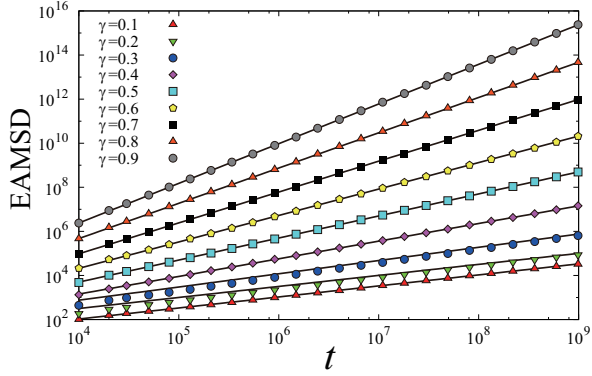}
  \caption{Ensemble-averaged mean square displacements
    ($\alpha=0.5)$. Symbols are the results of numerical simulations for
    different $\gamma$ with theoretical lines. There are no
    fitting parameters in the theoretical lines.
    We set the PDF of the trapping time as $w(t) = \alpha t^{-1-\alpha}$ $(t\geq 1)$ in all the
    numerical simulations. Thus, the jump length PDF is given by $l(x) =
    \alpha/2\gamma |x|^{1+\alpha/\gamma} $ from
    Eq.~(\ref{e.pdf-jump-length}), and  $\langle l^2 \rangle =
    \alpha/(\alpha-2\gamma)$ for $2\gamma < \alpha$.}
  \label{msd_alpha=0.5}
\end{figure}

Here, we derive the asymptotic behavior of the moments of position $x_t$
for $t\rightarrow \infty$ using the Fourier-Laplace transform
$\hat{P}(k,s)$.
The Laplace transform of $\langle x_t \rangle$, denoted by $\langle x_s \rangle$, is given by
\begin{equation}
\langle x_s \rangle = -
  \left. i\frac{\partial {\hat{P}} (k,s)}{\partial k}\right|_{k=0}
  =0, 
  \end{equation}
  which means there is no drift, $\langle x_t \rangle=0$.
  Similarly, the Laplace
  transform of the second moment, i.e., the ensemble-averaged MSD
  (EAMSD), is given by
\begin{eqnarray}
  \langle x^2_s \rangle &=&
  \left. -\frac{\partial^2 {\hat{P}}(k,s)}{\partial k^2} \right|_{k=0}
  = -\frac {1}{s} \frac{{\hat{\psi}}''(0,s)}{1 - {\hat{w}}(s)}. 
\end{eqnarray}
Using the asymptotic behavior at $s\rightarrow 0$, we have 
\begin{equation}
  \label{e.eamsd-laplace}
  \langle x^2_s \rangle  \simeq
  \begin{cases}
    \dfrac{\Gamma(2\gamma-\alpha)}{|\Gamma (-\alpha)|}
    \dfrac{1}{s^{1+2\gamma}},
    &\quad (2\gamma > \alpha)\\[0.5cm]
    \dfrac {1}{|\Gamma(-\alpha)| s^{\alpha+1}} \log\left(\dfrac
    {1}{s}\right),
    &\quad (2\gamma = \alpha)\\[0.5cm]
    \dfrac {\left\langle t^{2\gamma} \right\rangle}{ c s^{\alpha+1}}.
    &\quad (2\gamma < \alpha)
  \end{cases}
\end{equation}
 The inverse Laplace
transform for $t\rightarrow \infty$ reads
\begin{equation}
  \langle x^2_t \rangle \simeq
  \begin{cases}
    \dfrac{\Gamma(2\gamma-\alpha)}{|\Gamma (-\alpha)| \Gamma(1+2\gamma)}  t^{2\gamma}, 
    &\quad (2\gamma > \alpha)\\[0.5cm]
    \dfrac {1}{|\Gamma(-\alpha)| \Gamma (1 + \alpha)} t^{\alpha} \log t,  
    &\quad (2\gamma = \alpha)\\[0.5cm]
    \dfrac {\left\langle l^2 \right\rangle}{c \Gamma (1 + \alpha)} t^{\alpha},
    &\quad (2\gamma < \alpha)
  \end{cases}
  \label{eamsd}
\end{equation}
where we used 
$\left\langle t^{2\gamma} \right\rangle = \left\langle l^{2}
\right\rangle = \int_{-\infty} ^\infty x^2 l(x) dx$. 
Figure~\ref{msd_alpha=0.5} shows the EAMSD for $\alpha=0.5$. Theory (\ref{eamsd}) is in 
excellent agreement with numerical simulations.

\begin{figure*}
  \includegraphics[width=.9\linewidth, angle=0]{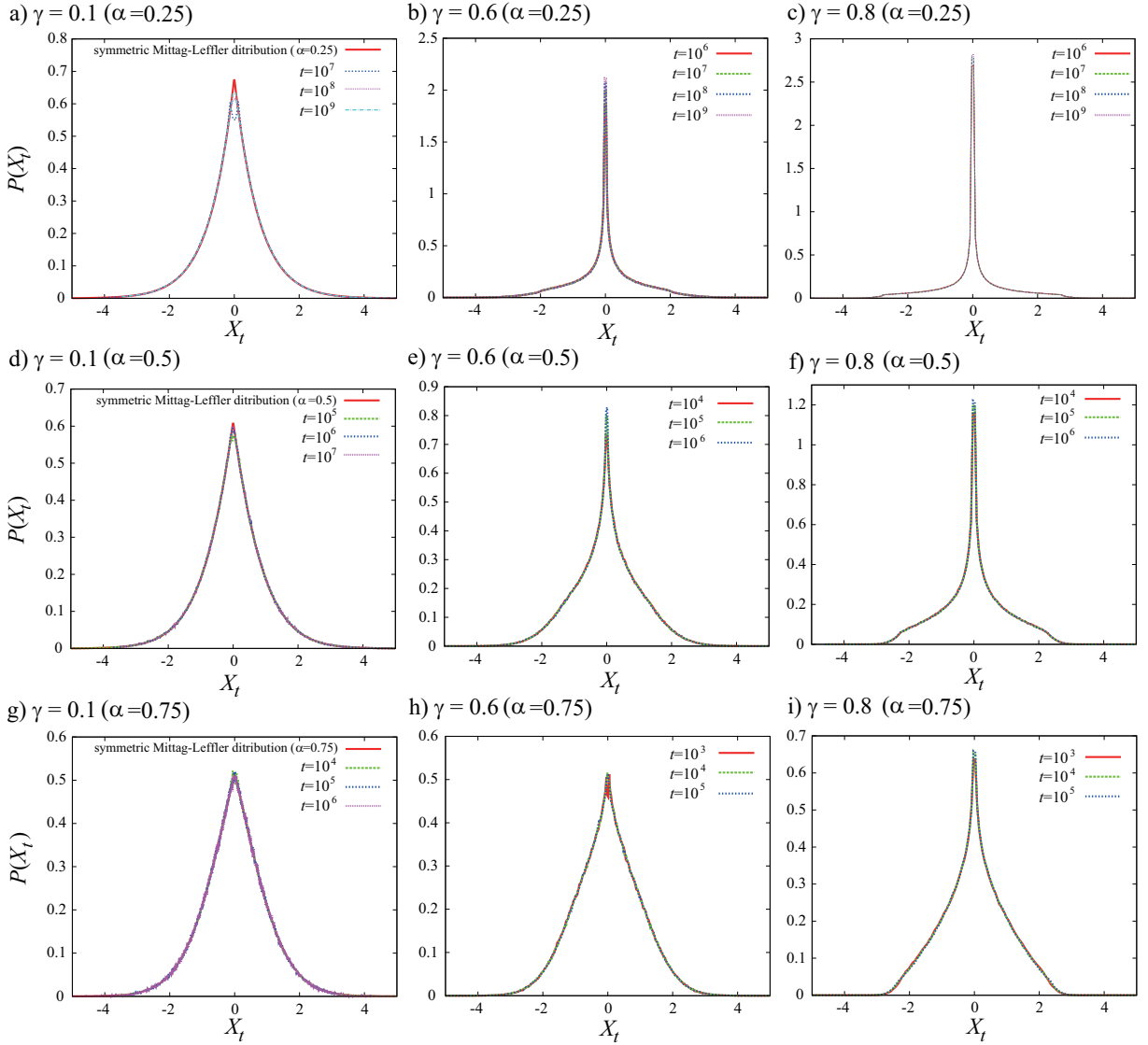}
  \caption{Probability density functions of a scaled position $X_t=x_t/\sqrt{\langle x_t^2\rangle}$ ($\alpha=0.25, 0.5$ and 0.75).  
  PDF $P(X_t)$ converges to a 
    non-trivial PDF as $t\rightarrow \infty$. (a), (d), and (g) the PDFs $P(X_t)$ converge to symmetric Mittag-Leffler distributions 
    ($\gamma=0.1$). For $2\gamma > \alpha$, the PDFs $P(X_t)$ converge to different distributions depending on $\alpha$ 
    as well as $\gamma$. 
    The PDF $w(t)$ used in the numerical simulation is the same as that in Fig.~\ref{msd_alpha=0.5}.}
  \label{pdf_alpha=0.5}
\end{figure*}

It is also possible to derive higher order moments in the following way. By
Eq.~(\ref{e.renewal-cctrw}), we have the
relation,
$\hat{P} (k,s)=  \hat{P} (k,s)\hat{\psi} (k,s) + C,$
where $C$ does not depend on $k$. Differentiating both sides $n$ times with respect to $k$, we
have
\begin{equation}
  \hat{P}^{(n)} (k,s)= \frac {1}{1 - \hat{\psi}(k,s)} \sum_{l=0}^{n-1} {}_nC_l \hat{P}^{(l)} (k,s) \hat{\psi}^{(n-l)}(k,s).
\end{equation}
From Eq.~(\ref{e.joint-prob-fourier-laplace}), we have
  $\hat{\psi}^{(2n+1)}(0,s) = 0$. Accordingly, we obtain
  ${\hat{P}}^{(2n+1)}(0,s)=0$
and 
\begin{equation}
  \label{e.recursion-P}
  \hat{P}^{(2n)} (0,s)= \frac {1}{1 - \hat{w}(s)} \sum_{l=0}^{n-1} {}_{2n}C_{2l} \hat{P}^{(2l)} (0,s) \hat{\psi}^{(2n-2l)}(0,s),
\end{equation}
by induction.  Thus, $\langle x^{2n+1}_t \rangle =0$ and  
the leading order for the Laplace transform of 
$\langle x^{2n}_t \rangle$ is given by 
\begin{eqnarray}
  \label{e.moment}
  \langle x^{2n}_s \rangle   \simeq
  \begin{cases}
    {\displaystyle 
      \dfrac{M_n(\alpha,\gamma)}{s^{1+2n\gamma}},}
    &\quad (2\gamma > \alpha)\\[0.3cm]
    \dfrac{ (2n)! \{-\hat{\psi}''(0,s) \}^n}{2^ns [1 - {\hat{w}}(s)]^{n}}, 
    &\quad (2\gamma \leq \alpha)
  \end{cases}
\end{eqnarray}
where $M_n(\alpha,\gamma)$ is given by a recursion relation 
$M_n(\alpha,\gamma) =  \sum_{l=1}^{n} {}_{2n}C_{2l} M_{n-l}(\alpha,
\gamma) \Gamma (2 l\gamma - \alpha)/|\Gamma(-\alpha)|.$ 
The above equations (\ref{e.moment}) can be confirmed by
Eq.~(\ref{e.recursion-P}) and mathematical induction.  Therefore, the
asymptotic behavior for $s\rightarrow 0$ is given by
\begin{eqnarray}
  \langle x^{2n}_s \rangle \simeq
  \begin{cases}
    \dfrac{M_n(\alpha,\gamma)}{s^{1+2n\gamma}},
    &\quad (2\gamma > \alpha)\\[0.5cm]
    \dfrac {(2n)! \left\{\log\left(\frac {1}{s}\right)\right\}^n}{2^n|\Gamma(-\alpha)|^n s^{n\alpha+1}},
    &\quad (2\gamma = \alpha)\\[0.5cm]
    \dfrac {(2n)!\left\langle t^{2\gamma} \right\rangle^n}{ (2c)^n s^{n\alpha+1}},
    &\quad (2\gamma < \alpha)
  \end{cases}
\end{eqnarray}
and the inverse Laplace transform for $t\rightarrow \infty$ reads 
\begin{equation}
  \langle x^{2n}_t \rangle \simeq
  \begin{cases}
    \dfrac{M_n(\alpha,\gamma)}{\Gamma(1+2n\gamma)}  t^{2n\gamma}, 
    &\quad (2\gamma > \alpha)\\[0.5cm]
    \dfrac {(2n)!}{2^n |\Gamma(-\alpha)|^n \Gamma (1 + n\alpha)} t^{n\alpha} \{\log t\}^n, 
    &\quad (2\gamma = \alpha)\\[0.5cm]
    \dfrac {(2n)!\left\langle l^2 \right\rangle^n}{(2c)^n \Gamma (1 + n\alpha)} t^{n\alpha}.
    &\quad (2\gamma < \alpha)
  \end{cases}
\end{equation}
It follows that the distribution of a scaled position $x_t/ \sqrt{\langle x_t^2 \rangle}$
converges to a time-independent non-trivial distribution. 
In other words, $x_t/\sqrt{\langle x_t^2 \rangle}$ converges in distribution to $X_{\alpha,\gamma}$ as $t\rightarrow \infty$, where 
\begin{equation}
  \langle e^{ikX_{\alpha,\gamma}} \rangle = 
  \begin{cases}
    {\displaystyle \sum_{n=0}^{\infty} 
      \frac{(ik)^{2n} M_n(\alpha,\gamma)\Gamma(1+2\gamma)^n}{(2n)! M_1(\alpha, \gamma)^n\Gamma(1+2n\gamma)}},  
    &\quad (2\gamma > \alpha) \\[.5cm]
    {\displaystyle \sum_{n=0}^{\infty} \frac{(ik)^{2n}\Gamma(1+\alpha)^{n}}{2^n\Gamma(1+n\alpha)}}.
    &\quad (2\gamma \leq \alpha)
  \end{cases}
\end{equation}
We note that the distribution of the random variable $X_{\alpha,\gamma}$
for $2\gamma \leq \alpha$ is called a symmetric Mittag-Leffler distribution
of order $\alpha/2$ \cite{kasahara77,Miyaguchi2013}.
Figure~\ref{pdf_alpha=0.5} shows the PDFs of $X_{\alpha,\gamma}$ for
$\gamma=0.1, 0.6$, and 0.8 ($\alpha=0.25$, 0.5, and 0.75).  For $2\gamma <
\alpha$, the PDFs converge to the symmetric Mittag-Leffler distribution,
which does not depend on $\gamma$. However, the PDFs are different from the
symmetric Mittag-Leffler distribution and depend crucially on $\gamma$ when
$2\gamma > \alpha$.

\section {Distributional ergodicity of time-averaged mean square displacement}
\begin{figure*}
  \includegraphics[width=.7\linewidth, angle=0]{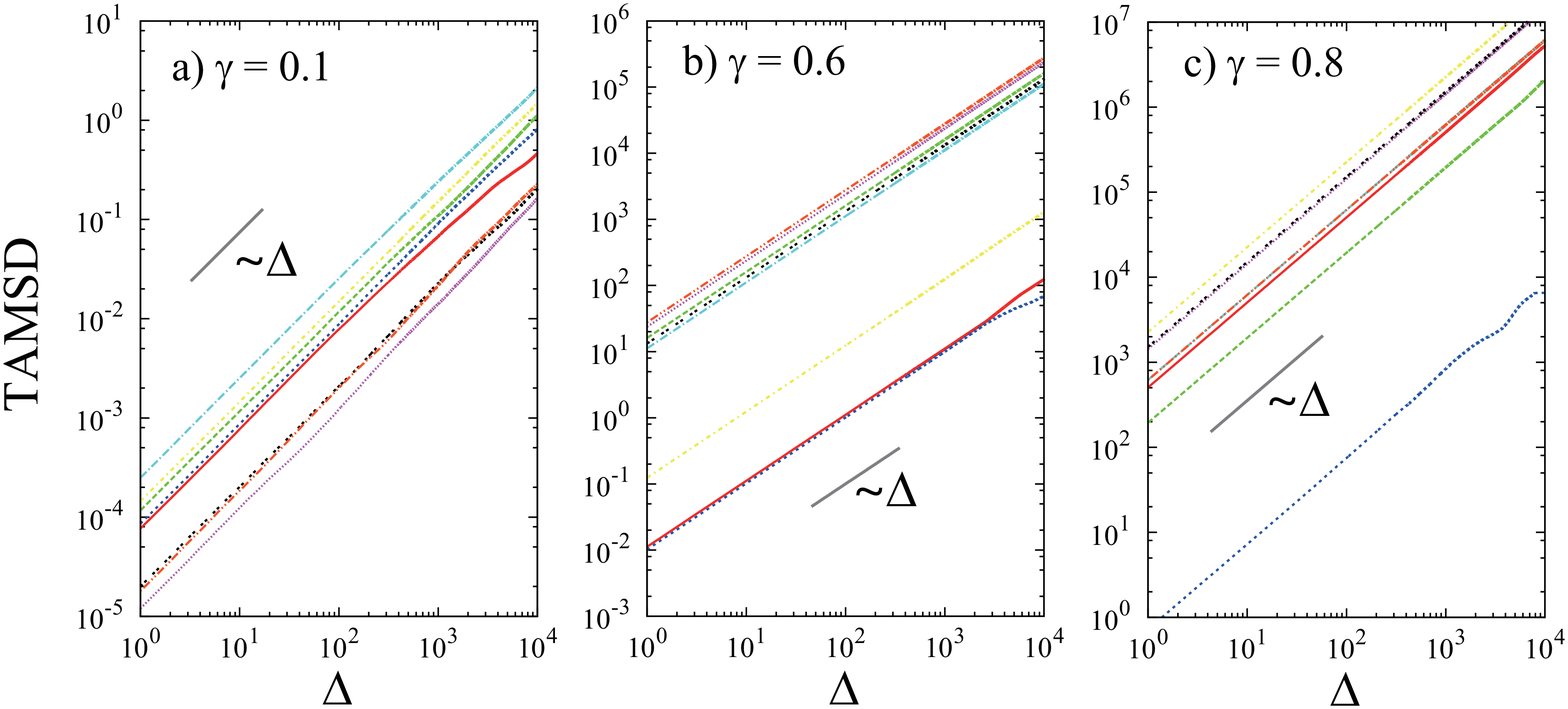}
  \caption{Time-averaged mean square displacements ($\alpha=0.5$ and $t=10^6$). TAMSDs for eight different realizations 
    are drawn in (a), (b), and (c) for $\gamma=0.1, 0.6$  and  $\gamma=0.8$, respectively.  Linear scalings are shown by the solid lines for 
    reference. The PDF $w(t)$ used in the numerical simulation is the same as that in Fig.~\ref{msd_alpha=0.5}.
  }
  \label{TAMSD_alpha=0.5}
\end{figure*}

Here, we investigate ergodic properties of
time-averaged MSD (TAMSD), defined by
\begin{equation}
  \overline{\delta^2 (\Delta; t)} \equiv \frac{1}{t-\Delta} \int_0^{t-\Delta} [x(t'+\Delta)-x(t')]^2 dt'.
\end{equation}
It has been known that TAMSD can be represented using the total number of
jumps \cite{Miyaguchi2011a, Miyaguchi2013}, i.e., $N_t$, and $h_k = \Delta
l_k^2 + 2\sum_{m=1}^{k-1}l_kl_m \theta (\Delta - t_k+t_m)$:
\begin{equation}
  \overline{\delta^2 (\Delta; t)} \simeq \frac{1}{t}\sum_{k=0}^{N_t} h_k\quad (t\rightarrow\infty),
  \label{tamsd_Nt}
\end{equation}
where $l_k$ is the $k$th jump, $t_k$ is the time $k$th jump occurs,
and $\theta(t)$ is a step function, defined by $\theta(t)=0$ for $t<0$ and
$t$ otherwise.
One can show that 
$\sum_{k=0}^{N_t} (h_k - \Delta l^2_k)/\sum_{k=0}^{N_t} l_k^2 \rightarrow 0$ as $t\rightarrow \infty$ (see Appendix A). It follows
\begin{equation}
  \label{tamsd_Nt2}
  \overline{\delta^2 (\Delta; t)} \simeq D_t \Delta 
  \quad (\Delta \ll t ~{\rm and}~t\rightarrow \infty),
\end{equation}
where $D_t = \sum_{k=0}^{N_t} l^2_k/t$. 
As shown in Fig.~\ref{TAMSD_alpha=0.5}, TAMSDs increase linearly with time $\Delta$  (normal diffusion),
while the diffusion coefficients show large fluctuations.

Now, we derive the PDF $P_2 (z,t)$ of $Z_t\equiv
\sum_{k=0}^{N_t} l^2_k$. 
We note that $l^2_k$  and $N_t$ are mutually correlated 
because both of them depend on the $k$th trapping time, and thus we cannot apply the 
method used in previous studies \cite{Miyaguchi2011a, Miyaguchi2013}. Instead, we use the fact 
 that $Z_t$ obeys a directed SEDLF with the joint probability
\begin{equation}
  \label{e.joint-prob-z}
  \psi_2 (z,t) = w(t) \delta(z-t^{2\gamma}). 
\end{equation}
Therefore, in the same way as Eq.~(\ref{e.renewal-cctrw}), we obtain
\begin{equation}
  \label{e.renewal-cctrw-z}
  {\hat{P}_2}(k,s) =
  \frac {1}{s}
  \frac{1-\hat{w}(s)}{1-{\hat{\psi}_2}(k,s)}, 
\end{equation}
with ${\hat{\psi}_2} (k,s) = \int_{0}^{\infty} e^{-st} e^{ikt^{2\gamma}} w(t) dt.$
Thus, the calculations of the moments $Z_t$ are almost
  parallel with the case of $x_t$. For example, we obtain the mean
  diffusion coefficient, $\langle D_t \rangle = \langle Z_t \rangle/t$, as
  follows:
\begin{align}
  \langle D_t \rangle
  &\simeq 
  \begin{cases}
    \frac{\Gamma(2\gamma-\alpha)}{|\Gamma (-\alpha)| \Gamma(1+2\gamma)} t^{2\gamma-1},
    & (2 \gamma > \alpha) \\[0.2cm]
    \frac{1}{|\Gamma (-\alpha)| \Gamma(1+\alpha)} t^{\alpha-1} \log t,
    & (2 \gamma = \alpha) \\[0.2cm]
    \frac{\left\langle l^2 \right\rangle }{c \Gamma(1+\alpha)} t^{\alpha-1},
    & (2 \gamma < \alpha)
  \end{cases}
\end{align}
for $t\rightarrow \infty$. For $2\gamma > 1$, the diffusion
coefficient enhances, otherwise it shows aging.  This enhancement of
diffusion coefficients is completely different from separable CTRWs
\cite{Lubelski2008, He2008, Miyaguchi2011, Miyaguchi2013} and correlated
CTRWs \cite{Tejedor2010}, both of which show only aging.

\begin{figure}
  \includegraphics[width=.9\linewidth, angle=0]{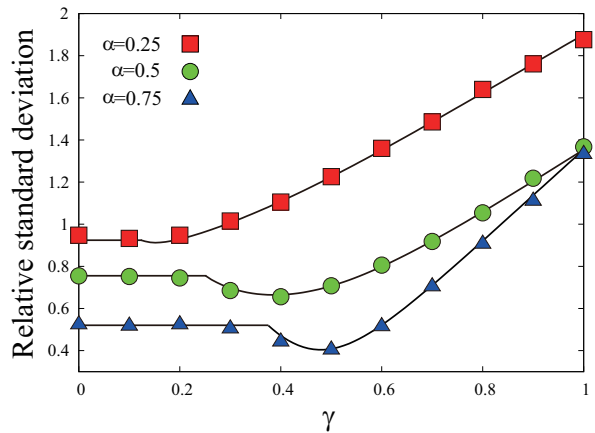}
  \caption{ Relative standard deviation of $D_t$ as a function of $\gamma$ ($\alpha=0.25, 0.5$, and 0.75). 
    Symbols are results of numerical simulations. We calculate $D_t$ by $\overline{\delta^2 (\Delta; t)}/\Delta$ in numerical simulations 
    with $\Delta = 10^3$. 
    Solid lines are theoretical curves (\ref{rsd_eq}).
    The PDF $w(t)$ used in the numerical simulation is the same as that in Fig.~\ref{msd_alpha=0.5}.}
  \label{RSD}
\end{figure}

Furthermore, the second moment of $D_t$ is given by
\begin{equation}
  \langle D^{2}_t \rangle \simeq
  \begin{cases}
    \frac{\Gamma(4\gamma-\alpha) |\Gamma (-\alpha)| + 2\Gamma(2\gamma-\alpha)^2}{\Gamma(4\gamma+1)|\Gamma(-\alpha)|^2}
    t^{4\gamma-2},
    & (2 \gamma > \alpha) \\[0.2cm]
    \frac {2}{\Gamma(2\alpha + 1) |\Gamma (-\alpha)|} \left(t^{\alpha} \log t\right)^{2},
    & (2 \gamma = \alpha) \\[0.2cm]
    \frac {2 \left\langle l^2 \right\rangle^2}{c^2 \Gamma (2 \alpha +1)} t^{2(\alpha-1)}.
    & (2 \gamma < \alpha)
  \end{cases}
\end{equation}
It follows that the relative standard deviation (RSD) of $D_t$, $\sigma_{\rm EB}
\equiv \sqrt{\langle D_t^2 \rangle - \langle D_t \rangle^2}/\langle D_t
\rangle $, which is an ergodicity breaking parameter \cite{He2008,
Miyaguchi2011, Miyaguchi2011a, Akimoto2011, Uneyama2012}, remains constant
as $t\rightarrow \infty$:
\begin{align}
  \sigma_{\rm EB}
  \simeq
  \begin{cases}
    \sqrt{
      \left[
      \frac{ \Gamma(4\gamma-\alpha) | \Gamma(-\alpha) |}{\Gamma(2\gamma-\alpha)^2} +2
      \right] \Phi (2\gamma) -1}
    &(2\gamma>\alpha)\\[.5cm]
    \sqrt{2\Phi (\alpha)-1} &(2\gamma \leq \alpha),
  \end{cases}
  \label{rsd_eq}
\end{align}
where $\Phi (x) \equiv {\Gamma(x+1)^2}/{\Gamma(2x+1)}$.  As shown in
Fig.~\ref{RSD}, the RSDs of $D_t$ depend on $\gamma$, and they are
different from that in CTRW when $2\gamma > \alpha$. Moreover, when
$2\gamma=1$, distributional behavior of diffusion coefficients of TAMSDs
appears intrinsically whereas the EAMSD is normal.

\begin{figure*}
  \includegraphics[width=.9\linewidth, angle=0]{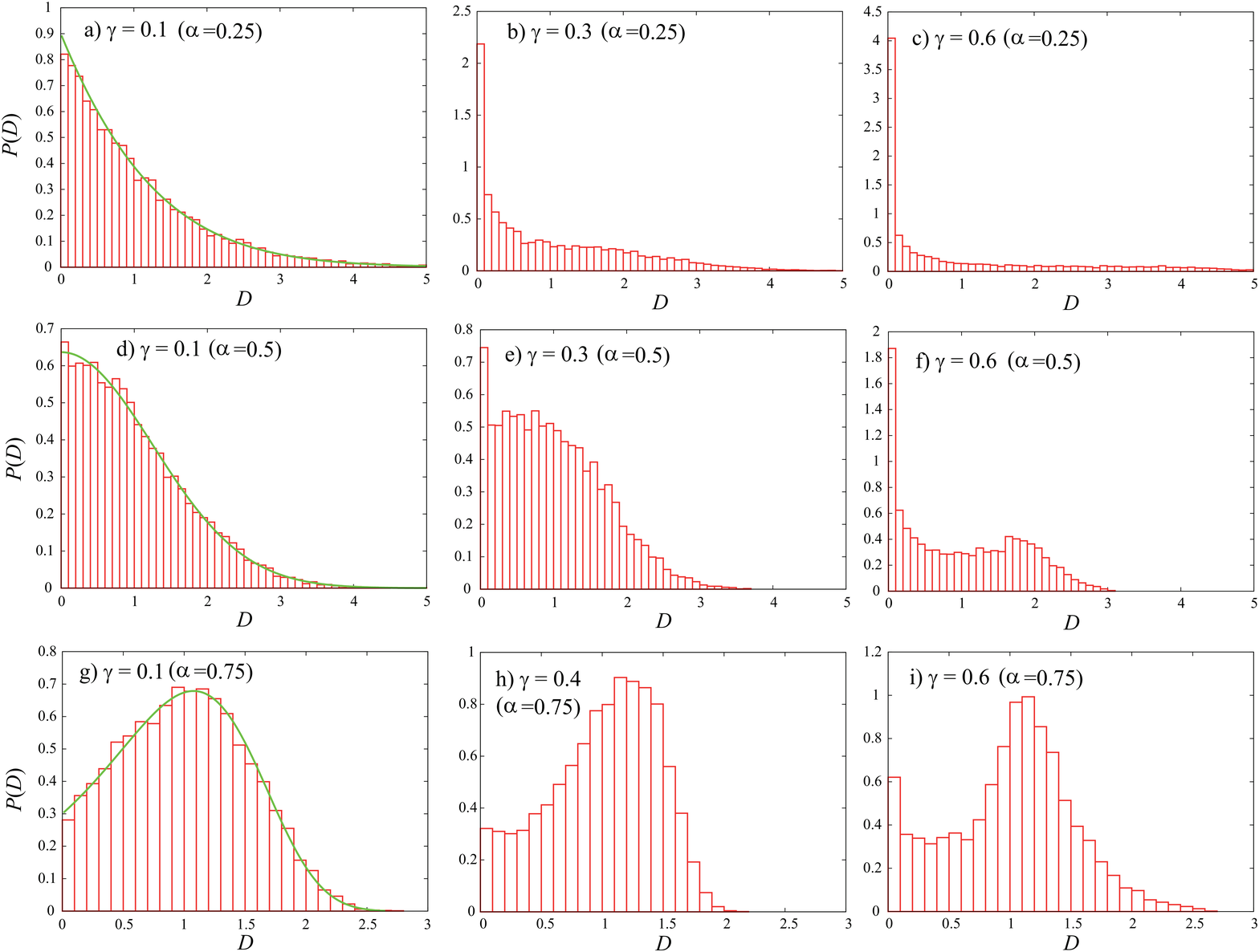}
  \caption{ Histograms of the normalized diffusion coefficients $D\equiv D_t/\langle D_t \rangle$ for different $\gamma=0.1$, 0.4, 0.6, and 0.8 ($\alpha=0.25$ 
  and 0.75). 
  $D_t$ is calculated in the same way as in Fig.~\ref{RSD}. The solid curves represent the Mittag-Leffler distribution. 
  The PDF $w(t)$ used in the numerical simulation is the same as that in Fig.~\ref{msd_alpha=0.5}.}
  \label{pdfd}
\end{figure*}

In a way similar to the calculation for $x_t$, we can obtain all the higher
moments of $D_t$. In particular, for $2\alpha < \gamma$, 
\begin{equation}
\langle D^n_t \rangle \simeq 
\frac{n!\left\langle l^2 \right\rangle^n }{c^n \Gamma(n+\alpha)} t^{n(\alpha-1)}.
\end{equation}
Therefore, the distribution of the scaled diffusion coefficient $D\equiv D_t/
\langle D_t \rangle$ converges to the Mittag-Leffler distribution, i.e., 
the Laplace transform of the random variable $D$ is given by
\begin{equation}
\langle e^{zD} \rangle = \sum_{n=0}^\infty \frac{\Gamma(1+\alpha)^nz^n}{\Gamma(1+n\alpha)}.
\end{equation}
Moreover, for $2\gamma > \alpha$, the distribution of $D$ also converges to a time-independent non-trivial
distribution as $t\to \infty$, indicating that the scaled diffusion
coefficient converges to a random variable (i.e., distributional ergodicity). 
PDFs of the normalized diffusion coefficient $D$ for different parameters are shown in 
Fig.~\ref{pdfd} by
numerical simulations. PDFs depend crucially on the coupling parameter
$\gamma$ for $2\gamma > \alpha$. We note that the PDF for $2\gamma <
\alpha$ is exactly the same as the Mittag-Leffler distribution of order
$\alpha$.

\section {Conclusion}

In conclusion, we have shown subdiffusion as well as superdiffusion in the
SEDLF using Laplace analysis.  By numerical simulations, we have
presented the asymptotic behaviors of the  PDF of the normalized
positions in the SEDLF.  This model (SEDLF) removes unphysical situations
in L\'evy flight such that the EAMSD always diverges.  In the SEDLF, we
have shown that TAMSDs increase linearly with time and the diffusion
coefficients converge in distribution (distributional ergodicity).
 Distributions of the diffusion coefficients depends not only on
the exponent $\alpha$ of the trapping-time distribution but also on the
coupling exponent $\gamma$ for $2\gamma > \alpha$, and thus are different
from those in separable CTRWs \cite{He2008} as well as random walks with
static disorder \cite{Miyaguchi2011}. Especially, in superdiffusive regime
$(\gamma > 0.5)$, the mean diffusion coefficient enhances according to the
increase of the measurement time.  

\appendix

\section {Derivation of Eq.~(\ref{tamsd_Nt2})}

Here, we derive Eq.~(\ref{tamsd_Nt2}).
For $\gamma \in (0, \alpha/2)$, because of $\langle l^2_k \rangle < \infty$, both terms
\begin{equation}
\frac {1}{n}\sum_{k=1}^{n} \Delta l_k^2 \quad \text{and} \quad
\frac {2}{n}\sum_{k=1}^{n} \sum_{m=1}^{k-1}l_kl_m \theta (\Delta - t_k+t_m)
\end{equation}
converge to their ensemble averages as $n \to \infty$ thanks to the law of
large numbers. Moreover, the first term is dominant over the second because
the ensemble average of the second term is 0 from $\left\langle l_k
\right\rangle = 0$. Thus, we obtain the approximation given by
Eq.~(\ref{tamsd_Nt2}).

For $\gamma \in (\alpha/2, \alpha)$, the first term diverges as $n \to
\infty$ because of $\left\langle l_k^2 \right\rangle = \infty$, while the second
term remains finite. Thus, in this case too, the first term is dominant and
Eq.~(\ref{tamsd_Nt2}) is valid.

For $\gamma \in (\alpha, 1)$, the both terms diverge, but still the same
approximation holds. From the generalized limit theorem for stable
distributions \cite{bouchaud90},  the first term scales as
$\sim n^{2\gamma / \alpha - 1}$, because a random variable $y = l_k^2$ is distributed 
according to PDF $f(y) \sim
1/y^{1 + \alpha / 2\gamma}$. On the other hand, the second term scales as
$\lesssim n^{\gamma/\alpha-1}$ because
\begin{align}
  \notag
  \sum_{k=1}^{n} \sum_{m=1}^{k-1} |l_kl_m \theta (\Delta - t_k+t_m)|
   < 
  \triangle^{2} \sum_{k=1}^{n} |l_k| 
  \lesssim
  \triangle^{2} n^{\gamma/\alpha},
\end{align}
where we used the generalized central limit theorem again for the scaling
of $\sum_{m}^{}|l_m|$ (Note that $|l_m|$ follows the PDF $p(l) \sim 1 /
l^{1 + \alpha/\gamma}$). We also used the facts that $\tau_k +  \ldots
+\tau_{m+1} \leq \triangle$ if $\theta (\Delta - t_k+t_m) > 0$, where $\tau_k\equiv 
t_k - t_{k-1}$.
Thus, $\sum_{m=1}^{k-1} |l_m| \theta (\Delta - t_k+t_m) = \sum_{m=1}^{k-1} \tau^\gamma_m \theta (\Delta - t_k+t_m) <
  \triangle^{2}$. 
  Finally, the ratio of the
second term against the first goes to zero, i.e., $n^{\gamma/\alpha-1} /n^{2\gamma /
  \alpha - 1} = n^{ - \gamma/\alpha} \to 0$ as $n \to \infty$.


\begin{thebibliography}{30}%
\makeatletter
\providecommand \@ifxundefined [1]{%
 \@ifx{#1\undefined}
}%
\providecommand \@ifnum [1]{%
 \ifnum #1\expandafter \@firstoftwo
 \else \expandafter \@secondoftwo
 \fi
}%
\providecommand \@ifx [1]{%
 \ifx #1\expandafter \@firstoftwo
 \else \expandafter \@secondoftwo
 \fi
}%
\providecommand \natexlab [1]{#1}%
%
\providecommand \bibnamefont  [1]{#1}%
\providecommand \bibfnamefont [1]{#1}%
\providecommand \citenamefont [1]{#1}%
\providecommand \href@noop [0]{\@secondoftwo}%
\providecommand \href [0]{\begingroup \@sanitize@url \@href}%
\providecommand \@href[1]{\@@startlink{#1}\@@href}%
\providecommand \@@href[1]{\endgroup#1\@@endlink}%
\providecommand \@sanitize@url [0]{\catcode `\\12\catcode `\$12\catcode
  `\&12\catcode `\#12\catcode `\^12\catcode `\_12\catcode `\%12\relax}%
\providecommand \@@startlink[1]{}%
\providecommand \@@endlink[0]{}%
\providecommand \url  [0]{\begingroup\@sanitize@url \@url }%
\providecommand \@url [1]{\endgroup\@href {#1}{\urlprefix }}%
\providecommand \urlprefix  [0]{URL }%
%
\providecommand \doibase [0]{http://dx.doi.org/}%
\providecommand \selectlanguage [0]{\@gobble}%
\providecommand \bibinfo  [0]{\@secondoftwo}%
\providecommand \bibfield  [0]{\@secondoftwo}%
%
\providecommand \BibitemOpen [0]{}%
%
%
%
\providecommand \BibitemShut  [1]{\csname bibitem#1\endcsname}%
\let\auto@bib@innerbib\@empty
\bibitem [{\citenamefont {Caspi}\ \emph {et~al.}(2000)\citenamefont {Caspi},
  \citenamefont {Granek},\ and\ \citenamefont {Elbaum}}]{Caspi2000}%
  \BibitemOpen
  \bibfield  {author} {\bibinfo {author} {\bibfnamefont {A.}~\bibnamefont
  {Caspi}}, \bibinfo {author} {\bibfnamefont {R.}~\bibnamefont {Granek}}, \
  and\ \bibinfo {author} {\bibfnamefont {M.}~\bibnamefont {Elbaum}},\ }\href
  {\doibase 10.1103/PhysRevLett.85.5655} {\bibfield  {journal} {\bibinfo
  {journal} {Phys. Rev. Lett.}\ }\textbf {\bibinfo {volume} {85}},\ \bibinfo
  {pages} {5655} (\bibinfo {year} {2000})}\BibitemShut {NoStop}%
\bibitem [{\citenamefont {Golding}\ and\ \citenamefont
  {Cox}(2006)}]{Golding2006}%
  \BibitemOpen
  \bibfield  {author} {\bibinfo {author} {\bibfnamefont {I.}~\bibnamefont
  {Golding}}\ and\ \bibinfo {author} {\bibfnamefont {E.~C.}\ \bibnamefont
  {Cox}},\ }\href@noop {} {\bibfield  {journal} {\bibinfo  {journal} {Phys.
  Rev. Lett.}\ }\textbf {\bibinfo {volume} {96}},\ \bibinfo {pages} {098102}
  (\bibinfo {year} {2006})}\BibitemShut {NoStop}%
\bibitem [{\citenamefont {Gran\'{e}li}\ \emph {et~al.}(2006)\citenamefont
  {Gran\'{e}li}, \citenamefont {Yeykal}, \citenamefont {Robertson},\ and\
  \citenamefont {Greene}}]{Graneli2006}%
  \BibitemOpen
  \bibfield  {author} {\bibinfo {author} {\bibfnamefont {A.}~\bibnamefont
  {Gran\'{e}li}}, \bibinfo {author} {\bibfnamefont {C.~C.}\ \bibnamefont
  {Yeykal}}, \bibinfo {author} {\bibfnamefont {R.~B.}\ \bibnamefont
  {Robertson}}, \ and\ \bibinfo {author} {\bibfnamefont {E.~C.}\ \bibnamefont
  {Greene}},\ }\href@noop {} {\bibfield  {journal} {\bibinfo  {journal} {Proc.
  Natl. Acad. Sci. USA}\ }\textbf {\bibinfo {volume} {103}},\ \bibinfo {pages}
  {1221} (\bibinfo {year} {2006})}\BibitemShut {NoStop}%
\bibitem [{\citenamefont {Weigel}\ \emph {et~al.}(2011)\citenamefont {Weigel},
  \citenamefont {Simon}, \citenamefont {Tamkun},\ and\ \citenamefont
  {Krapf}}]{Weigel2011}%
  \BibitemOpen
  \bibfield  {author} {\bibinfo {author} {\bibfnamefont {A.}~\bibnamefont
  {Weigel}}, \bibinfo {author} {\bibfnamefont {B.}~\bibnamefont {Simon}},
  \bibinfo {author} {\bibfnamefont {M.}~\bibnamefont {Tamkun}}, \ and\ \bibinfo
  {author} {\bibfnamefont {D.}~\bibnamefont {Krapf}},\ }\href@noop {}
  {\bibfield  {journal} {\bibinfo  {journal} {Proc. Natl. Acad. Sci. USA}\
  }\textbf {\bibinfo {volume} {108}},\ \bibinfo {pages} {6438} (\bibinfo {year}
  {2011})}\BibitemShut {NoStop}%
\bibitem [{\citenamefont {Jeon}\ \emph {et~al.}(2011)\citenamefont {Jeon},
  \citenamefont {Tejedor}, \citenamefont {Burov}, \citenamefont {Barkai},
  \citenamefont {Selhuber-Unkel}, \citenamefont {Berg-S\o{}rensen},
  \citenamefont {Oddershede},\ and\ \citenamefont {Metzler}}]{Jeon2011}%
  \BibitemOpen
  \bibfield  {author} {\bibinfo {author} {\bibfnamefont {J.-H.}\ \bibnamefont
  {Jeon}}, \bibinfo {author} {\bibfnamefont {V.}~\bibnamefont {Tejedor}},
  \bibinfo {author} {\bibfnamefont {S.}~\bibnamefont {Burov}}, \bibinfo
  {author} {\bibfnamefont {E.}~\bibnamefont {Barkai}}, \bibinfo {author}
  {\bibfnamefont {C.}~\bibnamefont {Selhuber-Unkel}}, \bibinfo {author}
  {\bibfnamefont {K.}~\bibnamefont {Berg-S\o{}rensen}}, \bibinfo {author}
  {\bibfnamefont {L.}~\bibnamefont {Oddershede}}, \ and\ \bibinfo {author}
  {\bibfnamefont {R.}~\bibnamefont {Metzler}},\ }\href {\doibase
  10.1103/PhysRevLett.106.048103} {\bibfield  {journal} {\bibinfo  {journal}
  {Phys. Rev. Lett.}\ }\textbf {\bibinfo {volume} {106}},\ \bibinfo {pages}
  {048103} (\bibinfo {year} {2011})}\BibitemShut {NoStop}%
\bibitem [{\citenamefont {Weber}\ \emph {et~al.}(2012)\citenamefont {Weber},
  \citenamefont {Spakowitz},\ and\ \citenamefont {Theriot}}]{Weber2012}%
  \BibitemOpen
  \bibfield  {author} {\bibinfo {author} {\bibfnamefont {S.~C.}\ \bibnamefont
  {Weber}}, \bibinfo {author} {\bibfnamefont {A.~J.}\ \bibnamefont
  {Spakowitz}}, \ and\ \bibinfo {author} {\bibfnamefont {J.~A.}\ \bibnamefont
  {Theriot}},\ }\href@noop {} {\bibfield  {journal} {\bibinfo  {journal} {Proc.
  Natl. Acad. Sci. USA}\ }\textbf {\bibinfo {volume} {109}},\ \bibinfo {pages}
  {7338} (\bibinfo {year} {2012})}\BibitemShut {NoStop}%
\bibitem [{\citenamefont {Zaid}\ \emph {et~al.}(2009)\citenamefont {Zaid},
  \citenamefont {Lomholt},\ and\ \citenamefont {Metzler}}]{Zaid2009}%
  \BibitemOpen
  \bibfield  {author} {\bibinfo {author} {\bibfnamefont {I.~M.}\ \bibnamefont
  {Zaid}}, \bibinfo {author} {\bibfnamefont {M.~A.}\ \bibnamefont {Lomholt}}, \
  and\ \bibinfo {author} {\bibfnamefont {R.}~\bibnamefont {Metzler}},\
  }\href@noop {} {\bibfield  {journal} {\bibinfo  {journal} {Biophys. J.}\
  }\textbf {\bibinfo {volume} {97}},\ \bibinfo {pages} {710} (\bibinfo {year}
  {2009})}\BibitemShut {NoStop}%
\bibitem [{\citenamefont {Metzler}\ and\ \citenamefont
  {Klafter}(2000)}]{metzler00}%
  \BibitemOpen
  \bibfield  {author} {\bibinfo {author} {\bibfnamefont {R.}~\bibnamefont
  {Metzler}}\ and\ \bibinfo {author} {\bibfnamefont {J.}~\bibnamefont
  {Klafter}},\ }\href {\doibase 10.1016/S0370-1573(00)00070-3} {\bibfield
  {journal} {\bibinfo  {journal} {Phys. Rep.}\ }\textbf {\bibinfo {volume}
  {339}},\ \bibinfo {pages} {1} (\bibinfo {year} {2000})}\BibitemShut {NoStop}%
\bibitem [{\citenamefont {Saxton}(2007)}]{saxton07}%
  \BibitemOpen
  \bibfield  {author} {\bibinfo {author} {\bibfnamefont {M.~J.}\ \bibnamefont
  {Saxton}},\ }\href {\doibase 10.1529/biophysj.106.092619} {\bibfield
  {journal} {\bibinfo  {journal} {Biophys. J.}\ }\textbf {\bibinfo {volume}
  {92}},\ \bibinfo {pages} {1178} (\bibinfo {year} {2007})}\BibitemShut
  {NoStop}%
\bibitem [{\citenamefont {Lubelski}\ \emph {et~al.}(2008)\citenamefont
  {Lubelski}, \citenamefont {Sokolov},\ and\ \citenamefont
  {Klafter}}]{Lubelski2008}%
  \BibitemOpen
  \bibfield  {author} {\bibinfo {author} {\bibfnamefont {A.}~\bibnamefont
  {Lubelski}}, \bibinfo {author} {\bibfnamefont {I.~M.}\ \bibnamefont
  {Sokolov}}, \ and\ \bibinfo {author} {\bibfnamefont {J.}~\bibnamefont
  {Klafter}},\ }\href@noop {} {\bibfield  {journal} {\bibinfo  {journal} {Phys.
  Rev. Lett.}\ }\textbf {\bibinfo {volume} {100}},\ \bibinfo {pages} {250602}
  (\bibinfo {year} {2008})}\BibitemShut {NoStop}%
\bibitem [{\citenamefont {He}\ \emph {et~al.}(2008)\citenamefont {He},
  \citenamefont {Burov}, \citenamefont {Metzler},\ and\ \citenamefont
  {Barkai}}]{He2008}%
  \BibitemOpen
  \bibfield  {author} {\bibinfo {author} {\bibfnamefont {Y.}~\bibnamefont
  {He}}, \bibinfo {author} {\bibfnamefont {S.}~\bibnamefont {Burov}}, \bibinfo
  {author} {\bibfnamefont {R.}~\bibnamefont {Metzler}}, \ and\ \bibinfo
  {author} {\bibfnamefont {E.}~\bibnamefont {Barkai}},\ }\href@noop {}
  {\bibfield  {journal} {\bibinfo  {journal} {Phys. Rev. Lett.}\ }\textbf
  {\bibinfo {volume} {101}},\ \bibinfo {pages} {058101} (\bibinfo {year}
  {2008})}\BibitemShut {NoStop}%
\bibitem [{\citenamefont {Miyaguchi}\ and\ \citenamefont
  {Akimoto}(2011{\natexlab{a}})}]{Miyaguchi2011a}%
  \BibitemOpen
  \bibfield  {author} {\bibinfo {author} {\bibfnamefont {T.}~\bibnamefont
  {Miyaguchi}}\ and\ \bibinfo {author} {\bibfnamefont {T.}~\bibnamefont
  {Akimoto}},\ }\href@noop {} {\bibfield  {journal} {\bibinfo  {journal} {Phys.
  Rev. E}\ }\textbf {\bibinfo {volume} {83}},\ \bibinfo {pages} {062101}
  (\bibinfo {year} {2011}{\natexlab{a}})}\BibitemShut {NoStop}%
\bibitem [{\citenamefont {Meroz}\ \emph {et~al.}(2013)\citenamefont {Meroz},
  \citenamefont {Sokolov},\ and\ \citenamefont {Klafter}}]{Meroz2013}%
  \BibitemOpen
  \bibfield  {author} {\bibinfo {author} {\bibfnamefont {Y.}~\bibnamefont
  {Meroz}}, \bibinfo {author} {\bibfnamefont {I.~M.}\ \bibnamefont {Sokolov}},
  \ and\ \bibinfo {author} {\bibfnamefont {J.}~\bibnamefont {Klafter}},\ }\href
  {\doibase 10.1103/PhysRevLett.110.090601} {\bibfield  {journal} {\bibinfo
  {journal} {Phys. Rev. Lett.}\ }\textbf {\bibinfo {volume} {110}},\ \bibinfo
  {pages} {090601} (\bibinfo {year} {2013})}\BibitemShut {NoStop}%
\bibitem [{\citenamefont {Miyaguchi}\ and\ \citenamefont
  {Akimoto}(2011{\natexlab{b}})}]{Miyaguchi2011}%
  \BibitemOpen
  \bibfield  {author} {\bibinfo {author} {\bibfnamefont {T.}~\bibnamefont
  {Miyaguchi}}\ and\ \bibinfo {author} {\bibfnamefont {T.}~\bibnamefont
  {Akimoto}},\ }\href@noop {} {\bibfield  {journal} {\bibinfo  {journal} {Phys.
  Rev. E}\ }\textbf {\bibinfo {volume} {83}},\ \bibinfo {pages} {031926}
  (\bibinfo {year} {2011}{\natexlab{b}})}\BibitemShut {NoStop}%
\bibitem [{\citenamefont {Aaronson}(1997)}]{Aaronson1997}%
  \BibitemOpen
  \bibfield  {author} {\bibinfo {author} {\bibfnamefont {J.}~\bibnamefont
  {Aaronson}},\ }\href@noop {} {\emph {\bibinfo {title} {An Introduction to
  Infinite Ergodic Theory}}}\ (\bibinfo  {publisher} {American Mathematical
  Society},\ \bibinfo {address} {Providence},\ \bibinfo {year}
  {1997})\BibitemShut {NoStop}%
\bibitem [{\citenamefont {Akimoto}\ and\ \citenamefont
  {Miyaguchi}(2010)}]{Akimoto2010}%
  \BibitemOpen
  \bibfield  {author} {\bibinfo {author} {\bibfnamefont {T.}~\bibnamefont
  {Akimoto}}\ and\ \bibinfo {author} {\bibfnamefont {T.}~\bibnamefont
  {Miyaguchi}},\ }\href@noop {} {\bibfield  {journal} {\bibinfo  {journal}
  {Phys. Rev. E}\ }\textbf {\bibinfo {volume} {82}},\ \bibinfo {pages}
  {030102(R)} (\bibinfo {year} {2010})}\BibitemShut {NoStop}%
\bibitem [{\citenamefont {Meerschaert}\ and\ \citenamefont
  {Scalas}(2006)}]{Meerschaert2006}%
  \BibitemOpen
  \bibfield  {author} {\bibinfo {author} {\bibfnamefont {M.~M.}\ \bibnamefont
  {Meerschaert}}\ and\ \bibinfo {author} {\bibfnamefont {E.}~\bibnamefont
  {Scalas}},\ }\href@noop {} {\bibfield  {journal} {\bibinfo  {journal}
  {Physica A}\ }\textbf {\bibinfo {volume} {370}},\ \bibinfo {pages} {114}
  (\bibinfo {year} {2006})}\BibitemShut {NoStop}%
\bibitem [{\citenamefont {Helmstetter}\ and\ \citenamefont
  {Sornette}(2002)}]{Helmstetter2002}%
  \BibitemOpen
  \bibfield  {author} {\bibinfo {author} {\bibfnamefont {A.}~\bibnamefont
  {Helmstetter}}\ and\ \bibinfo {author} {\bibfnamefont {D.}~\bibnamefont
  {Sornette}},\ }\href@noop {} {\bibfield  {journal} {\bibinfo  {journal}
  {Phys. Rev. E}\ }\textbf {\bibinfo {volume} {66}},\ \bibinfo {pages} {061104}
  (\bibinfo {year} {2002})}\BibitemShut {NoStop}%
\bibitem [{\citenamefont {Klafter}\ \emph {et~al.}(1987)\citenamefont
  {Klafter}, \citenamefont {Blumen},\ and\ \citenamefont
  {Shlesinger}}]{Klafter1987}%
  \BibitemOpen
  \bibfield  {author} {\bibinfo {author} {\bibfnamefont {J.}~\bibnamefont
  {Klafter}}, \bibinfo {author} {\bibfnamefont {A.}~\bibnamefont {Blumen}}, \
  and\ \bibinfo {author} {\bibfnamefont {M.~F.}\ \bibnamefont {Shlesinger}},\
  }\href {\doibase 10.1103/PhysRevA.35.3081} {\bibfield  {journal} {\bibinfo
  {journal} {Phys. Rev. A}\ }\textbf {\bibinfo {volume} {35}},\ \bibinfo
  {pages} {3081} (\bibinfo {year} {1987})}\BibitemShut {NoStop}%
\bibitem [{\citenamefont {Magdziarz}\ \emph {et~al.}(2012)\citenamefont
  {Magdziarz}, \citenamefont {Szczotka},\ and\ \citenamefont
  {{\.Z}ebrowski}}]{Magdziarz2012}%
  \BibitemOpen
  \bibfield  {author} {\bibinfo {author} {\bibfnamefont {M.}~\bibnamefont
  {Magdziarz}}, \bibinfo {author} {\bibfnamefont {W.}~\bibnamefont {Szczotka}},
  \ and\ \bibinfo {author} {\bibfnamefont {P.}~\bibnamefont {{\.Z}ebrowski}},\
  }\href@noop {} {\bibfield  {journal} {\bibinfo  {journal} {J. Stat. Phys.}\
  }\textbf {\bibinfo {volume} {147}},\ \bibinfo {pages} {74} (\bibinfo {year}
  {2012})}\BibitemShut {NoStop}%
\bibitem [{\citenamefont {Liu}\ and\ \citenamefont {Bao}(2013)}]{Liu2013}%
  \BibitemOpen
  \bibfield  {author} {\bibinfo {author} {\bibfnamefont {J.}~\bibnamefont
  {Liu}}\ and\ \bibinfo {author} {\bibfnamefont {J.-D.}\ \bibnamefont {Bao}},\
  }\href@noop {} {\bibfield  {journal} {\bibinfo  {journal} {Physica A}\
  }\textbf {\bibinfo {volume} {392}},\ \bibinfo {pages} {612} (\bibinfo {year}
  {2013})}\BibitemShut {NoStop}%
\bibitem [{\citenamefont {Shlesinger}\ \emph {et~al.}(1982)\citenamefont
  {Shlesinger}, \citenamefont {Klafter},\ and\ \citenamefont
  {Wong}}]{Shlesinger1982}%
  \BibitemOpen
  \bibfield  {author} {\bibinfo {author} {\bibfnamefont {M.}~\bibnamefont
  {Shlesinger}}, \bibinfo {author} {\bibfnamefont {J.}~\bibnamefont {Klafter}},
  \ and\ \bibinfo {author} {\bibfnamefont {Y.}~\bibnamefont {Wong}},\
  }\href@noop {} {\bibfield  {journal} {\bibinfo  {journal} {J. Stat. Phys.}\
  }\textbf {\bibinfo {volume} {27}},\ \bibinfo {pages} {499} (\bibinfo {year}
  {1982})}\BibitemShut {NoStop}%
\bibitem [{\citenamefont {Bouchaud}\ and\ \citenamefont
  {Georges}(1990)}]{bouchaud90}%
  \BibitemOpen
  \bibfield  {author} {\bibinfo {author} {\bibfnamefont {J.}~\bibnamefont
  {Bouchaud}}\ and\ \bibinfo {author} {\bibfnamefont {A.}~\bibnamefont
  {Georges}},\ }\href {\doibase 10.1016/0370-1573(90)90099-N} {\bibfield
  {journal} {\bibinfo  {journal} {Phys. Rep.}\ }\textbf {\bibinfo {volume}
  {195}},\ \bibinfo {pages} {127} (\bibinfo {year} {1990})}\BibitemShut
  {NoStop}%
\bibitem [{\citenamefont {Scher}\ and\ \citenamefont
  {Montroll}(1975)}]{Scher1975}%
  \BibitemOpen
  \bibfield  {author} {\bibinfo {author} {\bibfnamefont {H.}~\bibnamefont
  {Scher}}\ and\ \bibinfo {author} {\bibfnamefont {E.~W.}\ \bibnamefont
  {Montroll}},\ }\href@noop {} {\bibfield  {journal} {\bibinfo  {journal}
  {Phys. Rev. B}\ }\textbf {\bibinfo {volume} {12}},\ \bibinfo {pages} {2455}
  (\bibinfo {year} {1975})}\BibitemShut {NoStop}%
\bibitem [{\citenamefont {Cox}(1962)}]{Cox}%
  \BibitemOpen
  \bibfield  {author} {\bibinfo {author} {\bibfnamefont {D.~R.}\ \bibnamefont
  {Cox}},\ }\href@noop {} {\emph {\bibinfo {title} {Renewal theory}}}\
  (\bibinfo  {publisher} {Methuen},\ \bibinfo {address} {London},\ \bibinfo
  {year} {1962})\BibitemShut {NoStop}%
\bibitem [{\citenamefont {Kasahara}(1977)}]{kasahara77}%
  \BibitemOpen
  \bibfield  {author} {\bibinfo {author} {\bibfnamefont {Y.}~\bibnamefont
  {Kasahara}},\ }\href@noop {} {\bibfield  {journal} {\bibinfo  {journal}
  {Publ. RIMS, Kyoto Univ.}\ }\textbf {\bibinfo {volume} {12}},\ \bibinfo
  {pages} {801} (\bibinfo {year} {1977})}\BibitemShut {NoStop}%
\bibitem [{\citenamefont {Miyaguchi}\ and\ \citenamefont
  {Akimoto}(2013)}]{Miyaguchi2013}%
  \BibitemOpen
  \bibfield  {author} {\bibinfo {author} {\bibfnamefont {T.}~\bibnamefont
  {Miyaguchi}}\ and\ \bibinfo {author} {\bibfnamefont {T.}~\bibnamefont
  {Akimoto}},\ }\href {\doibase 10.1103/PhysRevE.87.032130} {\bibfield
  {journal} {\bibinfo  {journal} {Phys. Rev. E}\ }\textbf {\bibinfo {volume}
  {87}},\ \bibinfo {pages} {032130} (\bibinfo {year} {2013})}\BibitemShut
  {NoStop}%
\bibitem [{\citenamefont {Tejedor}\ and\ \citenamefont
  {Metzler}(2010)}]{Tejedor2010}%
  \BibitemOpen
  \bibfield  {author} {\bibinfo {author} {\bibfnamefont {V.}~\bibnamefont
  {Tejedor}}\ and\ \bibinfo {author} {\bibfnamefont {R.}~\bibnamefont
  {Metzler}},\ }\href@noop {} {\bibfield  {journal} {\bibinfo  {journal} {J.
  Phys. A}\ }\textbf {\bibinfo {volume} {43}},\ \bibinfo {pages} {082002}
  (\bibinfo {year} {2010})}\BibitemShut {NoStop}%
\bibitem [{\citenamefont {Akimoto}\ \emph {et~al.}(2011)\citenamefont
  {Akimoto}, \citenamefont {Yamamoto}, \citenamefont {Yasuoka}, \citenamefont
  {Hirano},\ and\ \citenamefont {Yasui}}]{Akimoto2011}%
  \BibitemOpen
  \bibfield  {author} {\bibinfo {author} {\bibfnamefont {T.}~\bibnamefont
  {Akimoto}}, \bibinfo {author} {\bibfnamefont {E.}~\bibnamefont {Yamamoto}},
  \bibinfo {author} {\bibfnamefont {K.}~\bibnamefont {Yasuoka}}, \bibinfo
  {author} {\bibfnamefont {Y.}~\bibnamefont {Hirano}}, \ and\ \bibinfo {author}
  {\bibfnamefont {M.}~\bibnamefont {Yasui}},\ }\href {\doibase
  10.1103/PhysRevLett.107.178103} {\bibfield  {journal} {\bibinfo  {journal}
  {Phys. Rev. Lett.}\ }\textbf {\bibinfo {volume} {107}},\ \bibinfo {pages}
  {178103} (\bibinfo {year} {2011})}\BibitemShut {NoStop}%
\bibitem [{\citenamefont {Uneyama}\ \emph {et~al.}(2012)\citenamefont
  {Uneyama}, \citenamefont {Akimoto},\ and\ \citenamefont
  {Miyaguchi}}]{Uneyama2012}%
  \BibitemOpen
  \bibfield  {author} {\bibinfo {author} {\bibfnamefont {T.}~\bibnamefont
  {Uneyama}}, \bibinfo {author} {\bibfnamefont {T.}~\bibnamefont {Akimoto}}, \
  and\ \bibinfo {author} {\bibfnamefont {T.}~\bibnamefont {Miyaguchi}},\
  }\href@noop {} {\bibfield  {journal} {\bibinfo  {journal} {J. Chem. Phys.}\
  }\textbf {\bibinfo {volume} {137}},\ \bibinfo {pages} {114903} (\bibinfo
  {year} {2012})}\BibitemShut {NoStop}%
\end{thebibliography}
\end {document}